# A comment on the superconductivity of $Sr_2CuO_{3+}$

T.H. Geballe
Dept of Applied Physics and Materials Science, Stanford University, Stanford 94305, USA

M. Marezio
CRETA/CNRS, 38042 Grenoble cedex 9, France

*Abstract*
We have revisited data in the literature and find compelling reasons for believing that enhanced superconductivity occurs in $Sr_2CuO_{3+}$ when 25% or more of the oxygen sites in the $CuO_2$ layers are vacant, contrary to the almost universally accepted assumption that superconducting interactions in the high $T_c$ cuprates occur in stoichiometric $CuO_2$ layers.

*1. General Considerations*

Liu et al. [1] have reopened the intriguing question concerning the origin of superconductivity in cuprate superconductor $Sr_2CuO_{3+}$. We prefer the equivalent formula $Sr_2CuO_{4-1+}$ because the structure is that of the well studied $La_2CuO_4$ family of the high temperature cuprate superconductors, whose Ba-doped member was discovered to be superconducting by Bednorz and Mueller. Instead doping the $CuO_2$ layers by cation substitution in the adjacent rocksalt layers as was done by Bednorz and Mueller, the doping in the $Sr_2CuO_{4-1+}$ is accomplished by oxygen vacancies as expressed by the 4-1+ oxygen coefficient. This new doping mechanism is surprisingly effective because the resulting $T_c$s are more than double that obtained with optimum doping using cation substitutions. A significant difference in the superconducting behavior can be anticipated simply by noting the formal valence of the Cu cations. In optimally doped $(La_{1.84}Sr_{.16})CuO_4$ with $T_c \sim 40K$, the formal valence of Cu is 2.16+, whereas it is 2.8+ in $Sr_2CuO_{4-1+0.4}$ with $T_c = 95K$ [1, 2]. This difference alone portends large differences in the filling of the d bands. We will show below by a reexamination of data in the literature that there are compelling reasons for believing that the vacancies occur in the $CuO_2$ layers of the superconducting phase. Consequently, we find that the current understanding of superconductivity in the high $T_c$ cuprates which relies on full oxygen occupancy for the $CuO_2$ layers is inadequate. Our interest in this system was stimulated by the possibility that the oxygen vacancies themselves might be active pairing centers [3]. A tight binding calculation of the pd bands in the $CuO_2$ layer, carried out by Harrison [4] with 25% vacant oxygen sites shows that there are .13 electrons in the highest $e_g$ band, with an equal number of holes in another $e_g$ band, and no holes in the $t_g$ bands suggesting new pairing possibilities.

*2. Identifying the superconducting phase in multiphase samples*

Liu et al. [1] and the materials discoverers, Hiroi et al. [5], assumed that these vacancies were located in the rock salt blocks, that is in the $(SrO)_2$ double layers. This unquestionably reasonable assumption, however, is not based upon any direct experimental evidence that we are aware of. Below we cite compelling evidence that the contrary is true—and that the vacancies are located in the $CuO_2$ layers of the majority phase. In previous investigations definitive evidence was found for the oxygen vacancies to be located in the $CuO_2$ layers [2, 6-8]. However, all authors inferred that the less than full volume superconducting signals that were observed might originate from minority phases. Minority phases have been found in all methods of synthesis including that employed by Liu et al. [1].

The minority-phase-explanation gained credence, unjustified as we note below, from experiments by Scott et al. [9]. They used a high pressure synthesis method, similar to that described by Han et al. [2], in that $KClO_3$ was used as the agent for oxidizing the starting $Sr_2CuO_3$ material. By the use of a scanning squid microscope, Scott et al. found traces of a superconducting oxychloride, $Sr_3Cu_2O_5Cl$, to be responsible for the small zero-field-cooled superconducting signals that were observed. Although a direct comparison is not possible, these signals were much smaller than those reported by Han et al. [2]. Furthermore, the reported lattice constants of the majority phases in the two investigations were distinctly different.

Liu et al. [1] used $SrO_2$ in their high pressure synthesis and thus eliminated the possibility of the oxychloride contamination. In Fig 1 we compare the superconducting behavior of the Liu et al. samples [1] with that of Han et al. [2]. In spite of the remarkable similarity, Liu et al. dismissed the Han et al. results with the comment: "these (i.e. the Han samples) were multiphase mixtures and the superconducting volume fraction was very small, so it was difficult to determine the detailed structure and to identify the superconducting phase definitely". This statement is not supported by a comparison of the curves shown in Fig 1. $T_c$s, and the even more significantly the magnitudes of the Meissner (flux expulsion) signals in the field-cooled measurements, are remarkably similar. These considerations suggest that the signals originate from the same majority phase in both cases, contrary to the assignment by Liu et al., but consistent with further analysis of the data given below.

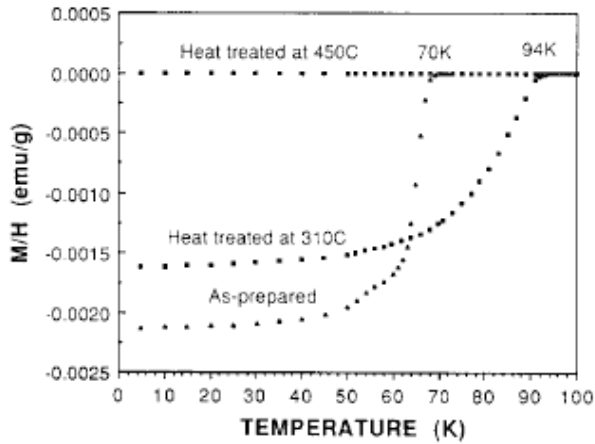

Diamagnetic susceptibility as a function of T for as prepared $Sr_2CuO_{3+\delta}$ and after different heat treatments. Han et al. Physica C 228, 129, 1994

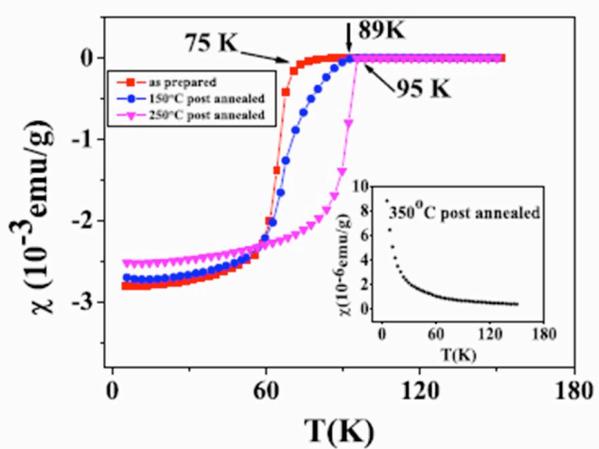

Temperature dependence of the dc magnetic susceptibility in the field cooling mode for as-prepared $Sr_2CuO_{3+\delta}$ samples and those after annealed at various T in $N_2$ atmosphere. Liu et al. Phys. Rev. 74, 100506(R), 2006

Liu et al. [1] have used x-ray, and electron diffraction as well as transmission electron imaging to characterize their samples. The majority phase (80%) found in their as-made sample is tetragonal I4/mmm with lattice constants a = 3.795, c = 12.507 Å. They report no significant change in that phase upon annealing to reach 95K. The Han et al. [2] samples have the same I4/mmm symmetry with lattice parameters a = 3.756, c = 12,521 Å and both sets of samples a clear evidence for an incommensurate modulated structure was given by electron diffraction.

A structural study on samples supplied by Hiroi et al. [5], the original discoverers, was performed by Shimakawa et al. [8]. These authors used the neutron diffraction powder technique and a simplified model of the structural modulation to carry out the refinements of the superconducting and non-superconducting phases. The refinements proved that the oxygen vacancies were essentially in the $CuO_2$ layers.

Subsequently, Wang et al. and Zhang et al. [6, 7] used the Han et al. samples [2] to study the evolution of the incommensurate-modulated majority phase structure upon annealing by transmission electron microscopy and diffraction and found changes in the modulation length as we discuss below. By analyzing the Cu cation displacement fields Zhang et al suggested a structural model for the oxygen-deficient $CuO_2$ layers. There would exist short and long Cu-Cu distances, with the oxygen atoms located only between the long Cu-Cu distances. This would correspond to $CuO_{1+}$ . This model is in qualitative agreement with that based on neutron diffraction powder data.

We would like to point out that the Liu et al. and Han et al. samples being compared here were made more than a decade apart in time, and in different laboratories using different methods of synthesis and with no apparent quantitative control of the oxygen concentration. The remarkable similarity of the superconductivity would not be likely if the superconducting signals were due to minority phases as Liu et al. postulate. Han et al. report the possible presence of minor phases of $SrO_2$, SrO, and CuO, whereas Liu et al. find the well-defined minority phase as described below. We have already noted that the majority phases in the Liu et al. and Han et al. samples are very similar.

The Liu et al. as-prepared sample consists of two phases with different modulations in the ab plane; the majority phase Fmmm is ~ 80% and the minority phase C2/m is less than 20%. Upon annealing to 250C the C2/m phase converts to a modulated Cmmm phase while they report that the majority phase remains unchanged. Liu et al assigned the 70K superconductivity in the as-made sample to the C2/m phase, and the 95K superconductivity in the annealed sample to the Cmmm phase. They argue that since the majority Fmmm phase does not change upon annealing, it cannot be responsible for the increased $T_c$. In so doing they disregard the work of Wang et al. [6] who examined the same Fmmm phase and found that upon annealing there is a change in the average incommensurate modulation wavelength from 4.8 to 5.2 in units of the primitive $\sqrt{2}a_p$ parameter, as well as a change in the oxygen K-edge fine structure.

Liu et al. [1] assigned the ~15% Meissner signals they observe upon field-cooling to the minority phase that occupies about 15% of their sample by volume. The field-cooled susceptibilities (Fig 1) are measured in low fields of 10 Oe and 20 Oe, respectively, for the Han et al. [2] and Liu et al. [1] samples. Usually, flux trapping and pinning is expected to preclude full flux expulsion in field-cooled experiments, particularly in multiphase samples. A relevant example where field-cooled (Meissner) signals are compared with zero-field-cooled (shielding) signals in a closely related cuprate, $Sr_{.9}La_{.1}CuO_2$, is given by Kim et al. [10]. These authors found that the Meissner signal was roughly 20% of the full-shielding (zero-field). Therefore it is quite reasonable to assume that the majority phase, common to both the Han and Liu samples, is responsible for the superconducting behavior and that the sizeable reduction in the Meissner signal is due to flux trapping. It would be an unusual coincidence if the minority phases in the two sets of samples were able to expel a similar amount of the flux in the field-cooled measurements. For the Liu et al samples it would require a full expulsion of the flux.

Liu et al. make the reasonable assumption that the enhanced ordering of the oxygen vacancies plays an important role in the enhancement of $T_c$ as did several groups earlier [2, 6-8]. However, they assume without any experimental evidence that the vacancies are in the apical oxygen sites. If that were true there should be modulations perpendicular to the SrO layers. The Cu ions would shift along the c-axis as the coordination goes from octahedral to pyramidal and the Sr ions would shift substantially along the c-axis as well to compensate for the missing charge. However, neither Liu et al. [1] nor do the comprehensive investigations by Wang et al. [6], Zhang et al. [7], Shimakawa et al. [8] and Laffez et al. [11] find evidence for a component of the modulation the along the c axis.

On the other hand the lack of c-axis modulation is expected if the oxygen vacancies are located in the $CuO_2$ layers. Cu cations linked to the missing oxygen sites move in the ab plane as the two apical oxygen ions remains symmetrical and the modulation is observed in that plane in agreement with the observations. The Sr cations would not be affected by the oxygen vacancies in the $CuO_2$ layers. According to the Zhang et al. model [7] the oxygen ions in the $CuO_{1+}$ layers are almost fully ordered. The presence of the  oxygens should not cause any Sr displacement in the annealed sample exhibiting $T_c$ = 95 K as these extra  ions are ordered too. On the other hand, the displacement of the Sr cations should also be minimal in the structure of the as-prepared sample where the extra  oxygens are not fully ordered.

*3. Evidence for competitive ordering ?*

Two sets of $Sr_2CuO_{4-1+}$ samples have been studied by neutron diffraction by Shimakawa et al. [5]. The powder diffraction data also show quite clearly that the vacancies are located in the $CuO_2$ layers and not in the $(SrO)_2$ layers. One of these sets made at ambient pressure was not superconducting down to 4K, while the other, made by the usual high pressure route,  had poorer superconducting behavior than  that shown in Fig 1: an onset $T_c$ of 65 K and a Meissner signal of 6%. The poorer superconducting behavior has encouraged a belief that the superconductivity arises from a different phase. We believe that this is not justified because the Han et al. samples, (as well as those of Liu et al.), become non superconducting when heated above 350C. Moreover, Wang et al. [4] investigated the non-superconducting phase that had been heated to 450C, before decomposition, but with a small loss (1.87%) in weight. The structure remained the same with a slight change in the modulation wavelength that was also consistent with the neutron samples. The density of states was estimated from EELS (electron energy loss) measurements of the oxygen K-absorption.. For the as-made and optimally annealed samples the pre-edge structure was sharp as expected for a high density of states near the Fermi surface, while for the non-superconducting sample it was low. This suggests that the disappearance of $T_c$ is due to a competing non-conducting ordered state that becomes the ground state with only a small change in the $CuO_{2-1+}$ layers. Consequently the lack of superconductivity does not justify disregarding the neutron diffraction results that independently find that the O vacancies are located in the $CuO_2$ layers.

*4. Conclusions*

The cuprate system $Sr_2CuO_{3+}$ , discovered by Hiroi et al. [5] differs significantly from all the models that have been employed to describe cuprate superconductivity. For example, 1) for  = 0.4 which corresponds to the maximum $T_c$ of 95 K, the average copper valence is ~ 2.8+ whereas is 2.15+ for all the others; 2) The remarkable similarity of the superconducting behavior of Han et al. (2) with the recent work of Liu et al. (1) gives renewed credibility to the possibility that the superconductivity originates in $CuO_2$ layers in which about 30% of the oxygen sites are vacant; 3) There is some evidence for competitive ordered states that become ground states with small changes obtained by heat treatment and different methods of preparation; 4) These findings offer new opportunities for investigating the pairing mechanism(s); 5) The robust optimum $T_c$ of ~ 95K, twice as large as the $T_c$s that are found in comparable cuprates, suggests the possibility that oxygen vacancies themselves might be responsible for the enhanced superconductivity [4]. The quantitative relationship between the concentration of oxygen vacancies and $T_c$ remains to be investigated. 5) It is clear that further theoretical and experimental investigations using well characterized single crystals and films are needed.

*Acknowledgement:* We thank Walter A. Harrison for sharing his tight binding model with us prior to publication. We have also benefited from helpful conversations with J. F Mitchell concerning the neutron studies. One of us (THG) would like to acknowledge support from the Air Force Office of Scientfic Research.